\documentstyle[preprint,aps]{revtex}
\tightenlines
\begin{document}
\draft
\title{Ring vortex destabilization in supersaturated
 $^3$He-$^4$He liquid mixtures at low temperatures}
\author{D.M. Jezek$^1$, M. Pi$^2$, M. Barranco$^2$, R.J. Lombard$^3$,
and M. Guilleumas$^4$.}
\address{$^1$Departamento de F\'{\i}sica, Facultad de Ciencias Exactas
y Naturales, \\
Universidad de Buenos Aires, RA-1428 Buenos Aires, and \\
Consejo Nacional de Investigaciones Cient\'{\i}ficas y T\'ecnicas,
Argentine}
\address{$^2$Departament d'Estructura i Constituents de la Mat\`eria,
Facultat de F\'{\i}sica, \\
Universitat de Barcelona, E-08028 Barcelona, Spain}
\address{$^3$Division de Physique Th\'eorique, Institut de Physique
Nucl\'eaire \\ F-91406 Orsay, France}
\address{$^4$Dipartimento di Fisica, Universit\`a di Trento. 38050
Povo, Italy}

\date{\today}

\maketitle

\begin{abstract}

The effect of ring vortices on the
destabilization of supersaturated liquid helium mixtures at very low
temperatures is investigated as a function of pressure.
We have found that large ring vortices trigger the segregation of
$^3$He at smaller values of the concentration than
small ring vortices. Our calculations indicate that the existence of
ring vortices in the mixture is a possible mechanism to understand the
rather small degree of critical supersaturation found in the
experiments.

\end{abstract}


\narrowtext
\section*{}

\section{Introduction}

The study of supersaturated liquid helium mixtures at very low
temperatures ($T$) has received a renewed experimental interest
\cite{Sa91,Mi91,Sa92,Ma92,Mi94}. The main goals of
these experiments have been, on the one hand,  to determine
the crossover from the classical thermal regime, to the
quantum regime in the nucleation process of the pure $^3$He phase
from the $^3$He-$^4$He mixture, and on the other hand, to study
the mixture in the metastable supersaturated state, whose existence
has been established long ago \cite{La69,Wa69}.

One of the more interesting observations of recent experiments is
the rather small degree of supersaturation achieved in them,
of the order of 1 \% in \cite{Mi91,Ma92,Mi94} (see also \cite{Ch97}),
and below 0.5 \% in \cite{Sa91,Sa92,Sa94}. Similar supersaturations
were found in \cite{La69,Wa69}, but at that time no systematic studies
of the metastable phase were undertaken.

An extrapolation of the measured $^3$He chemical potential excess
$\Delta \mu_3$ along the coexistence line
yielded $\partial \Delta \mu_3 / \partial x \geq 0$ up to values
of the $^3$He concentration $x$ $>$ 16 \% \cite{Se69},
 indicating that the mixture
could be in the metastable, homogeneous phase up to that concentration.
That belief seemed further sustained by the calculations of Lifshitz
{\it et al} in their classical paper about the kinetics of the
nucleation process in liquid helium mixtures \cite{Li78}, and by
microscopic \cite{Kr93} and density functional \cite{Gu95} estimates
of the spinodal line in the $P-x$ plane at zero temperature.

Clearly, there is nothing {\it a priori} contradictory between
the results of \cite{Kr93,Gu95} and the small critical
supersaturation degree $\Delta x_{cr}$
found experimentally. Likely, the segregation of $^3$He
from the mixture occurs before reaching the spinodal line. What is
really intriguing is that nucleation calculations
 \cite{Ma92,Li78,Je95}, which incorporate
the basic physics to consider $^3$He
droplets as nucleation seeds for phase separation, have
yielded results that are around one order of magnitude larger than
experiment. Other  calculations have
put the emphasis in the diffusion mechanism of the quantum decay
of the metastable mixture  and in the quantum-to-thermal
crossover temperature $T^*$ \cite{Bu93},
 taking as an input the experimental
$\Delta x_{cr}$. Yet, $T^*$ comes out a factor of four larger than
experiment.

It is then quite natural to look for another physical mechanism
as a possible explanation for the disagreement. One such mechanism
is heterogeneous nucleation. However, the system is extremely
pure, and the experimental cells are covered with $^4$He, avoiding
nucleation on the walls. Since at the low temperatures involved in
these experiments, the solubility of $^3$He into $^4$He is limited
and $^4$He is still superfluid, one is thus left with an obvious
candidate,
namely heterogeneous nucleation on vortices whose core is filled
with $^3$He \cite{Do91}. This possibility was already mentioned in
\cite{Mi94}, independently put forward and worked out in same detail
in
\cite{Je95}, and further developed in \cite{Je97}, as well as
in \cite{Bu96}.

The basis of the model employed in \cite{Je95} is that a $^4$He
vortex line described by a hollow core model (HCM) is gradually
filled with $^3$He as $x$ increases. In the metastable region,
these lines become unstable above a certain $\Delta x$ value, and the
resulting $\Delta x_{cr}$  turns out to be in the right order of
magnitude (around 1.5 \%), especially as compared with the
experimental results
of \cite{Mi91,Ma92}. The HCM results have been confirmed by a more
elaborated density functional calculation \cite{Je97}.

The purpose of the present work is to complete the above scenario,
studying the effect that  a more realistic vortex geometry, namely
a ring vortex, may have on the quantitative results, and also to
compare them with the recent experimental results of \cite{Ch97}.
As in previous works \cite{Je95,Je97,Bu96}, the presence of vortices
is taken for granted. Indeed, it has been  recognized for a long time
that `any container of superfluid $^4$He, treated in conventional
fashion, will be permeated {\it ab initio} by numerous quantized
vortices stabilized by surface pinning' \cite{Aw84}.
The interesting problem of vorticity nucleation
in liquid $^4$He, and the effect of $^3$He impurities on it  has
been addressed in detail in the past (see for example
\cite{Ra69,Wi78,Na85,Mu85,Av93}).

This work is organized as follows. We present in Section II the model
we have used to describe the evolution of a ring vortex as a function
of pressure and $^3$He concentration. The numerical results are
collected in Section III. Although our main interest is in the
metastable region in the pressure-concentration
($P-x$) plane at very low temperatures,
we have also paid some attention to the stable region, in view
that previous studies have not dealt with appreciable $^3$He
concentrations. A discussion is presented in Section IV, and some
technical details are given in an Appendix.

\section{Ring vortex energetics}

To carry out the program we have outlined in the introduction, we
first determine the kinetic energy of the ring vortex (RV), for which
we assume a  hollow core model,  i.e., the $^4$He density
is zero within the vortex core, and equal to the bulk value at
the given pressure elsewhere. Following \cite{Sc73,Fe76},
 we consider
the liquid to be incompressible ($\nabla \cdot \vec{v} \approx 0$), so
that one can introduce a velocity vector potential $\vec{A}(\vec{r}\,)$
such that $ \vec{v} = \nabla \times \vec{A}$. The vector potential
is fixed if the vorticity $\vec{\omega} =\nabla \times \vec{v}$
is specified:

\begin{equation}
\vec{\omega} = \nabla \times ( \nabla \times \vec{A}\,) \,\,\, .
\label{eq1}
\end{equation}

We make use of the line-source approximation for the vorticity. For a
ring vortex with radius of curvature $R$, circulation number $n$ = 1, 2,
..., and quantum circulation
$k_0 = n\, h/m_4$ located on the z = 0 plane, it means that

\begin{equation}
\vec{\omega}(\vec{r}) = k_0 \,\delta (r-R)\, \delta (z) \,
\hat{\phi}     \, \, ,
\label{eq2}
\end{equation}
where ($r, z, \phi$) are the cylindrical coordinates and
$ \hat{\phi} = ( -sin \phi, cos \phi, 0)$ is a unit vector.
Unless explicitly indicated, we shall be considering
$n = 1$ as circulation number.
It is straightforward to show that the integral solution of Eq.
(\ref{eq1}) reads $\vec{A}(\vec{r}\,) = A_0 (r,z) \hat{\phi}$, with

\begin{equation}
A_0(r,z) = \frac{k_0}{4 \pi} R
\int^{2 \pi}_{0} \frac{ cos \phi\,' d\phi\,'}{\sqrt{r^2 + R^2 -
2 \,r R \,cos \phi\,' + z^2}}
\, \, .
\label{eq3}
\end{equation}
The velocity of the superfluid outside the RV can be obtained as

\begin{eqnarray}
v_r & = & - \frac{\partial A_0}{\partial z}
\nonumber
\\
& &
\label{eq4}
\\
v_z & = & \frac{1}{r} \frac{\partial}{\partial r} ( r A_0) \,\, ,
\nonumber
\end{eqnarray}
and the kinetic energy of the superfluid is evaluated as an
integral over the core surface ${\cal S}$ of the ring vortex:

\begin{equation}
E_{kin}  = - \frac{m_4}{2} \rho_4
\int_{\cal S} A_0(r,z)\, (v_z \hat{r} - v_r \hat{z})
\cdot \hat{n} \, d {\cal S} \, \, ,
\label{eq5}
\end{equation}
where the normal $\hat{n}$ is directed into the $^4$He fluid. The
boundary condition that no velocity line crosses the vortex core
imposes that $\vec{v} \cdot \hat{n}$ = 0 and yields

\begin{equation}
\hat{n}  = - \frac{v_z}{\mid\vec{v}\mid} \hat{r}
+ \frac{v_r}{\mid\vec{v}\mid} \hat{z}
\label{eq6}
\end{equation}
as well as the differential equation obeyed by the velocity lines:

\begin{equation}
 \frac{d r}{d z} =  \frac{v_r}{v_z} \,\, .
\label{eq7}
\end{equation}

Using Eq. (\ref{eq4}), it is easy to verify that solving Eq.
(\ref{eq7}) to obtain the velocity lines is equivalent to solving the
algebraic equation

\begin{equation}
 r A_0(r,z) = Constant \,\, .
\label{eq8}
\end{equation}
We have specified the cross section profile of the vortex
core, compatible
with the shape of the velocity lines, introducing a length $a$ such
that $(R - a)$ is the closest distance of the core surface to the
origin of coordinates:

\begin{equation}
 (R-a) \, A_0(R-a,0) = C_a \,\, .
\label{eq9}
\end{equation}
That fixes the constant in Eq. (\ref{eq8}) and permits one to obtain
the core profile. In the case of  large ring vortices with $R>>a$, for
which a circular torus is a good approximation, the length $a$
represents the radius of the circular cross section of the core.
We have
numerically checked that if $a/R<<1$, $A_0$ at the surface of the
core, and the kinetic energy $E_{kin}$ are indeed those of the hollow
RV  given for example in \cite{Sc73}:

\begin{eqnarray}
A_0 & = &   \frac{k_0}{2 \pi}  \left[ \ln \frac{8 R}{a} - 2  \right]
\nonumber
\\
& &
\label{eq10}
\\
E_{kin} & = &   \frac{m_4}{2}\, \rho_4 \,k^2_0 \,R
  \left[ \ln \frac{8 R}{a} - 2  \right]  \,\, .
\nonumber
\end{eqnarray}
We give in the Appendix the general expressions of $A_0$, $v_r$ and
$v_z$ which are needed to evaluate the kinetic energy, Eq.
(\ref{eq5}).

Figure \ref{fig1} shows several  cross sections of the core for
$R$ = 30 ${\rm \AA}$ and $a$ = 10, 15 and 20 ${\rm \AA}$,
respectively. One can see that these cross sections are very
non-circular, especially when the ratio $a/R$ is large.

It is now a simple task to write the energy gain caused by the presence
of a ring vortex in the homogeneous mixture at given $P$. As $x$
increases, the $^3$He atoms located at the surface of the vortex
migrate to the interior of the hollow core.
Within the HCM, one has \cite{Je95}

\begin{equation}
 E_v(R,a) = \sigma {\cal S} - \Delta \mu_3 \, \rho_3 \,{\cal V} +
 E_{kin} \,\, ,
\label{eq11}
\end{equation}
where ${\cal V}$ is the volume of the core, $\rho_3$ the $^3$He
atom density inside the vortex core,
$\sigma$ the surface tension of the mixture-pure $^3$He interface, and
$\Delta \mu_3$ is the $^3$He chemical potential excess, i.e.,
the difference between the chemical potential of $^3$He in the
metastable mixture, and in pure $^3$He. Consequently, $\Delta \mu_3$
has a sign. It is negative when the mixture is stable, and positive
when it is not.

The reliability of eq. (\ref{eq11}) to  describe $^4$He vortices
filled with $^3$He has been checked in \cite{Je97} for the case of
vortex lines using a density functional approach which takes into
account the surface diffuseness of both liquids at the interface, and
the kinetic energy of the $^4$He atoms that
penetrate into the vortex core. Altogether, both effects
have been  found to be small, the reason being that the core radii
of critical vortices
are large, of the order of 10 ${\rm \AA}$, and consequently, the sharp
surface approximation embodied in the HCM works well. For example, at
zero $T$ and $P$, the HCM yields $\Delta x_{cr}$ $\sim$ 1.6 \%
\cite{Je95}, whereas the density functional theory yields
$\Delta x_{cr}$ $\sim$ 1.8 \% \cite{Je97}.

The temperatures of interest here are below 150 mK
(\cite{Sa92,Ma92,Ch97}). Bearing in
mind that we are looking for a sizeable decrease in the value of
$\Delta x_{cr}$ with respect to that of \cite{Li78}, and
 that thermal effects are not expected to play an important role
because of the small $T$-dependence of $\sigma$ \cite{Sa96},
$\Delta \mu_3$ \cite{Se69} and $\rho_3$ \cite{Ke62} in that
temperature domain,   we shall carry out a $T$ = 0 calculation.

At given  $P$ and $^3$He concentration $x \equiv \rho_3/
(\rho_3 + \rho_4)$, where $\rho_4$ is the $^4$He atom
density, the thermodynamical characteristics of the liquid have
been described by a density functional that among other
properties, reproduces accurately
the maximum solubility line and surface tension of the interface as
 a function of $P$ \cite{Ba97}. We give in Fig.  \ref{fig2}
 $\Delta \mu_3$ as a function of $x$ for $P$ = 0, 5 and 10 atm.
The $\Delta
\mu_3$ curves intersect the x-axis at a concentration $x_s(P)$
that corresponds to the maximum solubility of $^3$He into $^4$He for
the given $P$. The slopes $\partial (\Delta \mu_3) / \partial
x \mid_{x_s}$ will be used in the next section to determine the
critical supersaturation of the mixture when ring vortices are
present, transforming the $\Delta \mu_3$ values used in Eq.
(\ref{eq11}) into $\Delta x$ values as

\begin{equation}
 \Delta x \approx \Delta \mu_3 \left[ \frac{\partial \Delta \mu_3}
{\partial x} \mid_{x_s} \right]^{-1}  \,\, .
\label{eq12}
\end{equation}
This is justified in view of the smallness of $\Delta x$. We have
obtained $\partial \Delta \mu_3 / \partial x \mid_{x_s}$ = 2.49,
1.97 and 1.85  K, and $x_s$= 6.40, 9.01 and 9.58 \% for
$P$= 0, 5 and 10 atm, respectively.

\section{Results}

\subsection{Stable phase-diagram region}

We have first considered the region in the  $P-x$ plane where
the homogeneous mixture is stable, i.e., $P\geq 0$ and $\Delta x
\equiv x - x_s < 0$, and have minimized the  energy
Eq. (\ref{eq11})  of a RV of radius of curvature
$R$ with respect to $a$. The value of $R$ is considered as an intrinsic
characteristic of the vortex that depends on the process of formation,
whereas the $a$ value may change with changing $x$, and its optimum
value for given $x$ has to be found through minimization.

As expected from our previous
works on vortex lines \cite{Je95,Je97}, in this case there is only
one extremum of $E_v (R,a)$ corresponding to a stable ring vortex.
The energy of the equilibrium configuration and the effective
radius  of the  vortex core $a_0$, defined as
$\pi a^2_0 \equiv {\cal S}$,
are drawn in Figs.  \ref{fig3}  and  \ref{fig4}  as functions of
$R$ for $P$ = 0 and 10 atm, and $\Delta x$ = -1, -3 and -5 \%.

\subsection{Metastable phase-diagram region}

When $\Delta x$ is positive, $E_v(R,a)$ at fixed $R$ has in general
two extrema, one maximum and one minimum: the ring vortex is no
longer stable and may trigger the demixing process of $^3$He if
it overcomes the barrier whose height $\Delta E_v$ is the energy
difference between
both extrema. It may happen either by thermal or by quantum
fluctuations. One such energy barrier is drawn in Fig.  \ref{fig5}
for an $R$ = 100 ${\rm \AA}$ ring vortex at $P$ = 0 and
$\Delta x$ = 1 \%.
We show in Fig.  \ref{fig6}  the barrier heights as a function of
$\Delta x$, for different $R$  and $P$ values.

The barrier height decreases with increasing $\Delta x$, and there
is a critical value $\Delta x_{cr}$ at which it disappears. This
defines a critical ring vortex that depends on the values of $R$
and $P$, and may freely expand since it is barrierless.

We display in Fig. \ref{fig10} the effective radius $a_0$ of the
critical vortex as a function of $R$ for $P$ = 0, 5 and 10 atm. We see
that in contradistinction with the stable case, the core radii are
large, and as in the vortex line case,
we expect the sharp surface approximation to work well.
The critical configurations corresponding to $R$ =
30 ${\rm \AA}$ and $P$ = 0, 5 and 10 atm are shown in Fig. \ref{fig7}.
The associated $\Delta x_{cr}$
values are 2.75, 1.82 and 1.37 \%, respectively. For $R$ = 100 ${\rm
\AA}$, these values are 1.67, 1.04 and 0.76 \%.

From the previous discussion, it is clear that one can define, for a
given $R$-vortex, a kind of ring vortex spinodal line in the
$P-x$ plane, lying in the metastable region, beyond which a liquid
helium mixture hosting $R$-vortices is absolutely unstable.
Such spinodal lines can be drawn combining the appropiated
$(P, \Delta x)$ values from Fig.  \ref{fig6}. We have represented
in Fig  \ref{fig8}  $\Delta x_{cr}$ as a function of $R$ for
$P$ = 0, 5 and 10 atm. From this figure one can see that fixed $P$,
large ring vortices have associated smaller values of $\Delta x_{cr}$.
It means that large ring vortices trigger the seggregation of $^3$He
at smaller supersaturations that small ring vortices.

\section{Discussion}

Very recently, Chavogets et al \cite{Ch97} have carried out a systematic
investigation of nucleation in superfluid $^3$He-$^4$He solutions
at $T$ = 125 mK as a function of $P$,  which is especially well
suited for a detailed comparison with the present calculations since
the temperature is kept constant at a value of 125 $\pm$ 5 mK.
However, these authors stress that
it cannot be excluded that the experimental results
are influenced by the way the phase transition is initiated, since
different methods are used to attain the initial metastable state.

We show in Fig.  \ref{fig9}  the experimental results of \cite{Ch97}
for $\Delta x_{cr}$ (dots) and the spinodal values (solid lines)
corresponding to ring
vortices of $R$ = 100, 200 and 500 ${\rm \AA}$. For $R$ = 30 ${\rm \AA}$
the $\Delta x_{cr}$ values go from $\sim$ 2.7 \% at $P$ = 0
to $\sim$ 1.5 \% at $P$ = 10 atm.
The calculated $\Delta x_{cr}$'s tend to decrease with $P$, but
still are a factor of two larger than the experimental ones even for
the most favorable case of large ring vortices. However, the
improvement over nucleation calculations that do not consider
the possible existence of vortices in the mixture is apparent
{\cite{Mi91,Ch97,Li78}.

The major discrepancy appears at pressures
close to zero atm. Chagovets et al \cite{Ch97} have found that
$\Delta x_{cr}$ has
a maximum of 1.5 \% around $P$ = 0.1 atm, and that $\Delta x_{cr}$
drops to $\sim$ 0.7 \% around $P$ = 0.02 atm. We have not found this
behavior. However, we would like to recall that
at pressures close to zero,
the emergence of a free liquid-vapor interface may help nucleate
the new phase from the $^3$He-enriched surface layer \cite{Ma92}.
That might explain the small $\Delta x_{cr}$ found at low pressures,
although the authors of \cite{Ch97} have discarded this
possibility \cite{Ch98}.

The agreement between theory and experiment can be dramatically
improved  considering the existence of ring vortices with
circulation number $n > 1$. These vortices are rare, since creating an
$R$-vortex with $n$ = 2 costs more energy than creating two
$R$-vortices with $n$ = 1. This is the reason why in classical
experiments where vortices are nucleated and dragged by moving ions
only $n$ = 1 vortices are detected \cite{Ra64}.
Still, it is interesting to look at their effect on $\Delta x_{cr}$.
Roughly, one should expect that large RV with $n$ = 2
originate $\Delta x_{cr}$ four times smaller than vortices with $n$ = 1
(see for example the discussion after Eq. (4) of \cite{Je95}), thus
bringing theory closer to experiment. This is indeed the case,
as it can be seen from the dashed lines in Fig.  \ref{fig9}  which
represent the spinodal values corresponding to ring
vortices with $n$ = 2 and $R$ = 30, 50, 75, and 100 ${\rm \AA}$.

The above comparisons have involved the calculated
spinodal $\Delta x_{cr}$ values, which  are an upper bound to those
one would obtain considering thermal
activation over the energy barrier (quantum tunneling through it is
only effective for temperatures below a few tens of mK, see
\cite{Sa92}). We have checked that, although possible, thermal
activation plays a minor role. Let us consider, as an example, the
case of the $R$ = 50, $n$ = 2 ring vortex at $T$ = 125 mK, $P$ = 10
atm, for which the spinodal value is $\Delta x_{cr}$ = 0.56 \%. Within
classical nucleation theory, the decay rate $\Gamma$, i.e., the number of
RV `passing' over the barrier per unit time and volume
due to thermal fluctuations  is given by

\begin{equation}
 \Gamma =  \Gamma_0 \, exp(- \Delta E_v/kT)   \,\,\, ,
\label{eq13}
\end{equation}
where the prefactor $\Gamma_0$ can be roughly estimated as

\begin{equation}
 \Gamma_0 =  \frac{kT}{h R^3}  \,\,\, .
\label{eq14}
\end{equation}

Solving the equation 1 = $ V_{exp}\, t_{exp}\, \Gamma$, taking as
experimental volume $V_{exp}$ and time $t_{exp}$ those of \cite{Ch97},
and relating the $\Delta E_v$ value to $\Delta x$, one gets
$\Delta x_{cr}$ = 0.54 \% for $\Delta E_v$= 10 K, and $\Delta x_{cr}$
= 0.55 \% for $\Delta E_v$ = 8 K, which are very close to the
spinodal values.  A similar conclusion was drawn for vortex lines
in \cite{Je95}.

Quantum decay is effective only below a crossover temperature
$T^* \sim$ 10 mK \cite{Sa92}. None of the calculations carried out
so far is able to reproduce $\Delta x_{cr}$ and
$T^*$ simultaneously, and the problem is still
open. In the case of vortex lines, $T^*$ has been calculated to be
$\sim$ 0.1 mK in \cite{Bu96}. Using a functional integral method
similar to that of \cite{Je97b}, we have obtained $T^* \sim$ 1 mK.
The smallness of this value stems from the fact that in
the calculations, the decay process is always found to take place very
close to the spinodal line.

\acknowledgements

It is a pleasure to thank Valerii Chagovets for useful discussions,
and the authors of Ref.  \cite{Ch97} for sending us their
experimental data.
This work has been supported in part by the DGICYT (Spain) Grant No.
PB95-1249, by the Generalitat de Catalunya GRQ94-1022 program,
and by the IN2P3-CICYT agreement. M.G. thanks the Ministerio de
Educaci\'on y Cultura (Spain) for financial support.

\appendix
\section*{}

In this Appendix we give the expressions and method we have used
to numerically obtain the velocity lines. From Eqs.
(\ref{eq3}-\ref{eq4}) one has:

\begin{equation}
v_r(r,z) = \frac{k_0\, R\, z}{\pi \left[(r+R)^2+z^2\right]^{3/2}}
\int^{\pi/2}_0 \frac{2 \,cos^2\phi - 1}{(1 - s \,\,cos^2\phi)^{3/2}}
\,d\phi      \,\,\, ,
\label{a1}
\end{equation}
where we have defined $s \equiv 4 r R/ \left[(r+R)^2+z^2\right]$.
The integral Eq. (\ref{a1}) is written in terms of hypergeometric
functions \cite{Gr80} yielding:

\begin{equation}
v_r(r,z) = \frac{k_0\, R\, z}{2 \left[(r+R)^2+z^2\right]^{3/2}}
\left[ F(3/2, 3/2; 2; s) - F(1/2, 3/2; 1; s)\right]     \,\, .
\label{a2}
\end{equation}
Similarly,

\begin{eqnarray}
v_z(r,z) & = & \frac{k_0\, R}{2\, r \left[(r+R)^2+z^2\right]^{1/2}}
\left[ F(3/2, 1/2; 2; s) - F(1/2, 1/2; 1; s)\right]
\nonumber
\\
& + & \frac{k_0\, R}{2 \left[(r+R)^2+z^2\right]^{3/2}}
\left[ (r+R)\, F(1/2, 3/2; 1; s) - (r+2R)\,F(3/2, 3/2; 2; s)\right.
\nonumber
\\
& + & \frac{3R}{2}\, F(5/2, 3/2; 3; s) ]
\label{a3}
\end{eqnarray}
and
\begin{equation}
A_0(r,z)   =   \frac{k_0\, R}{2 \left[(r+R)^2+z^2\right]^{1/2}}
\left[ F(3/2, 1/2; 2; s) - F(1/2, 1/2; 1; s)\right]  \,\, .
\label{a4}
\end{equation}

The hypergeometric functions entering Eqs. (\ref{a2}-\ref{a4}) can be
written in terms of the complete elliptic integrals {\bf E} and
{\bf K} \cite{Gr80} as follows:

\begin{equation}
F(1/2,1/2;1;s)= \frac{2}{\pi} {\bf K}(\sqrt{s})
\label{a5}
\end{equation}

\begin{equation}
F(1/2,3/2;1;s)  =  F(3/2,1/2;1;s)
 =   \frac{2}{\pi} \frac{{\bf E}(\sqrt{s})}{1-s}
\label{a6}
\end{equation}

\begin{equation}
F(3/2,1/2;2;s)   =   F(1/2,3/2;2;s)
  =    \frac{4}{\pi s} [{\bf K}(\sqrt{s})- {\bf E}(\sqrt{s}) ]
\label{a7}
\end{equation}

\begin{equation}
F(3/2,3/2;2;s)  =  \frac{4}{\pi s} \left[
\frac{{\bf E}(\sqrt{s})}{1-s} -{\bf K}(\sqrt{s}) \right]
\label{a8}
\end{equation}

\begin{equation}
F(5/2,3/2;3;s)  =  \frac{16}{3 \pi s^2} \left[
\frac{2-s}{1-s}{\bf E}(\sqrt{s}) -2 {\bf K}(\sqrt{s}) \right] \,\, .
\label{a9}
\end{equation}
To evaluate {\bf E} and {\bf K} we have used accurate
polynomial approximations \cite{AS}.

\begin{figure}
\caption{ Core cross sections corresponding to  ring vortices
of radius of curvature
$R$ = 30 ${\rm \AA}$ and $a$-values of 10, 15 and 20 ${\rm \AA}$.
The  vorticity line is indicated by a dot.
}
\label{fig1}
\end{figure}
\begin{figure}
\caption{ $^3$He chemical potential excess $[\mu_3(x) - \mu_3(pure)]$
(K) as a function of $x$ (\%) for $P$ = 0, 5 and 10 atm.
}
\label{fig2}
\end{figure}
\begin{figure}
\caption{ Bottom panel: Ring vortex energy $E_v$ (K) as a function
of $R$ $({\rm \AA})$ for $P$ = 0.
Top panel: same as bottom panel  for $P$ = 10 atm. Both panels
correspond to stable vortices, and the curves
represent the results for $\Delta x$ = -1 \% (solid line), -3
\% (dotted line) and -5 \% (dashed line).
 }
\label{fig3}
\end{figure}
\begin{figure}
\caption{ Bottom panel: Effective radius $a_0$ $({\rm \AA})$
of the vortex cross section as a
function of $R$ $({\rm \AA})$  for $P$ = 0.
Top panel: same as bottom panel for $P$ = 10 atm. Both panels
correspond  to stable vortices, and  the curves
represent the results for $\Delta x$ = -1 \% (solid line), -3
\% (dotted line) and -5 \% (dashed line).
 }
\label{fig4}
\end{figure}
\begin{figure}
\caption{ Ring vortex energy $E_v$ (K) as a function of $a$
(${\rm \AA}$) for an $R$ = 100 ${\rm \AA}$ ring vortex at
$P$ = 0 and $\Delta x$ = 1 \%.
}
\label{fig5}
\end{figure}
\begin{figure}
\caption{ Barrier heights $\Delta E_v$ (K) as a function of $\Delta x$
for $P$ = 0, 5 and 10 atm. From top to bottom, the panels
correspond to $R$ = 10, 30 and 50 ${\rm \AA}$.
}
\label{fig6}
\end{figure}
\begin{figure}
\caption{ Effective radius $a_0$ $({\rm \AA})$
of the critical vortex as a
function of $R$ $({\rm \AA})$  for $P$ = 0, 5 and 10 atm.
 }
\label{fig10}
\end{figure}
\begin{figure}
\caption{ Core cross sections of critical ring vortices
 corresponding to  $R$ = 30 ${\rm \AA}$ and $P$ = 0, 5 and 10 atm.
}
\label{fig7}
\end{figure}
\begin{figure}
\caption{ Critical supersaturation $\Delta x_{cr}$ (\%) as a function
of $R$ for $P$ = 0, 5 and 10 atm.
}
\label{fig8}
\end{figure}
\begin{figure}
\caption{ Critical supersaturation $\Delta x_{cr}$ (\%) as a function
of $P$ (atm). The dots are the experimental values of Ref. [8].
From top to bottom, the solid lines represent the spinodal values for
ring vortices with $R$ = 100, 200 and 500 ${\rm \AA}$ and circulation
number $n$ = 1, and the dashed lines represent the spinodal values for
ring vortices with $R$ = 30, 50, 75 and 100 ${\rm \AA}$ and circulation
number $n$ = 2.
}
\label{fig9}
\end{figure}
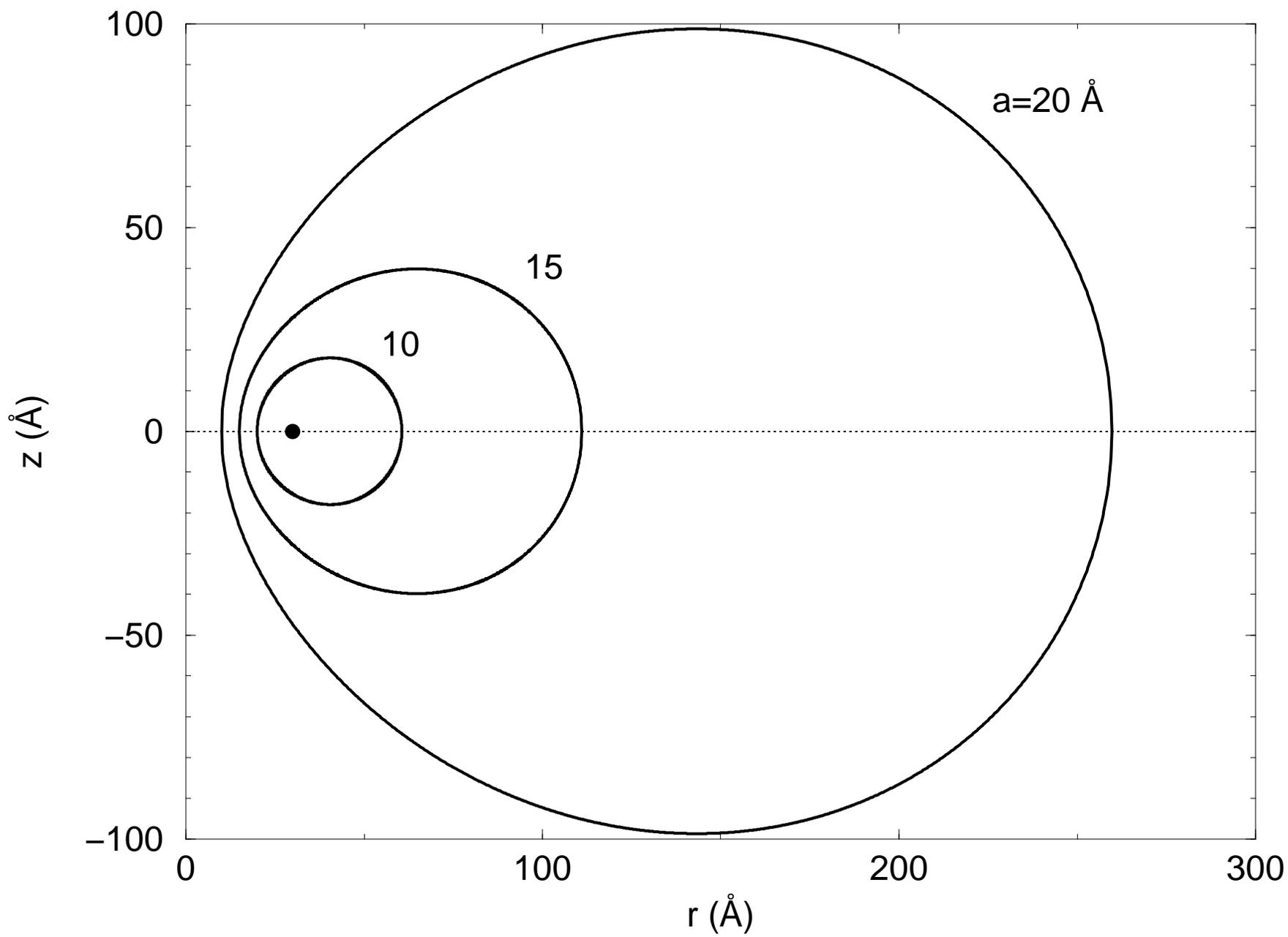
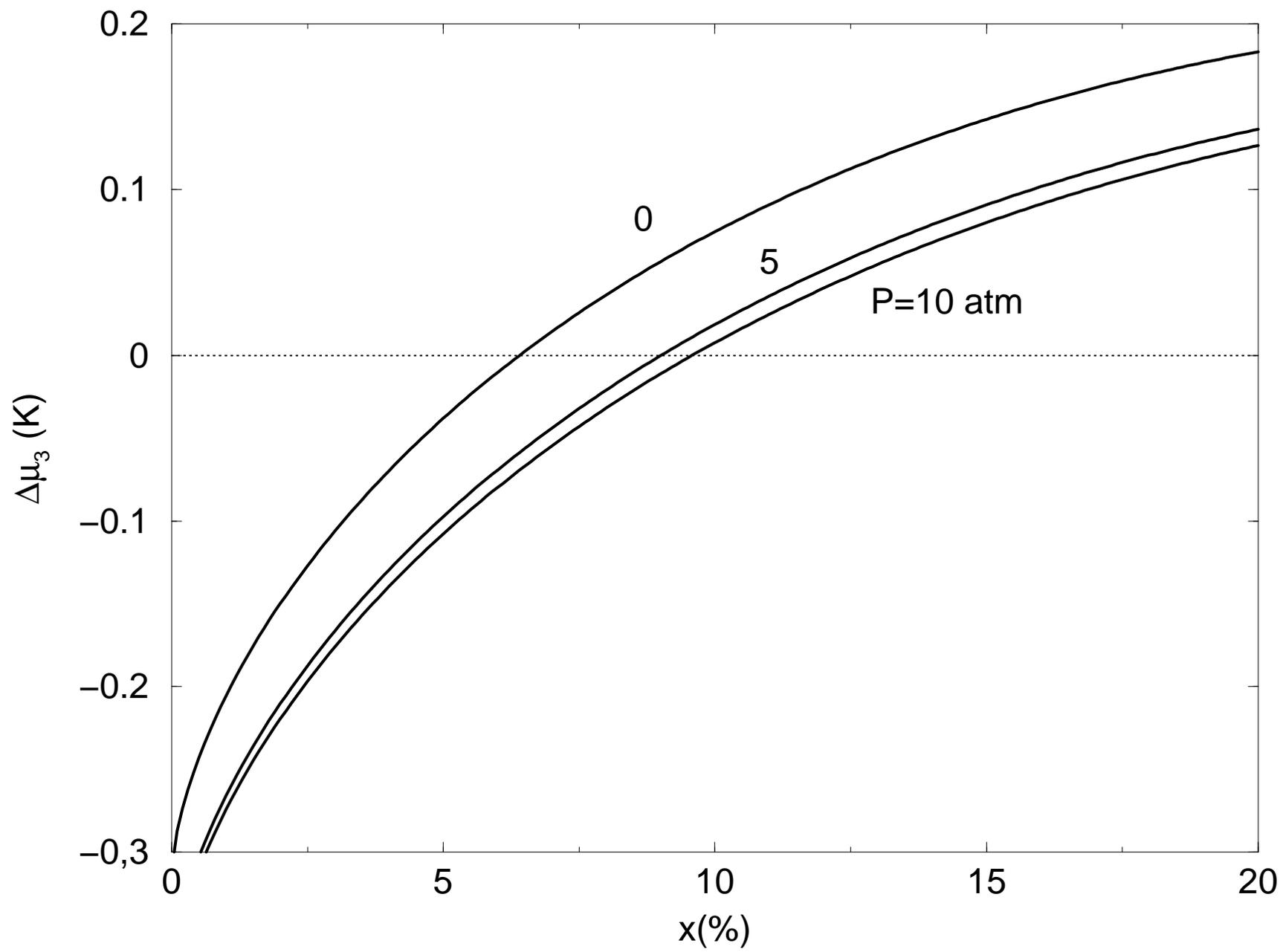
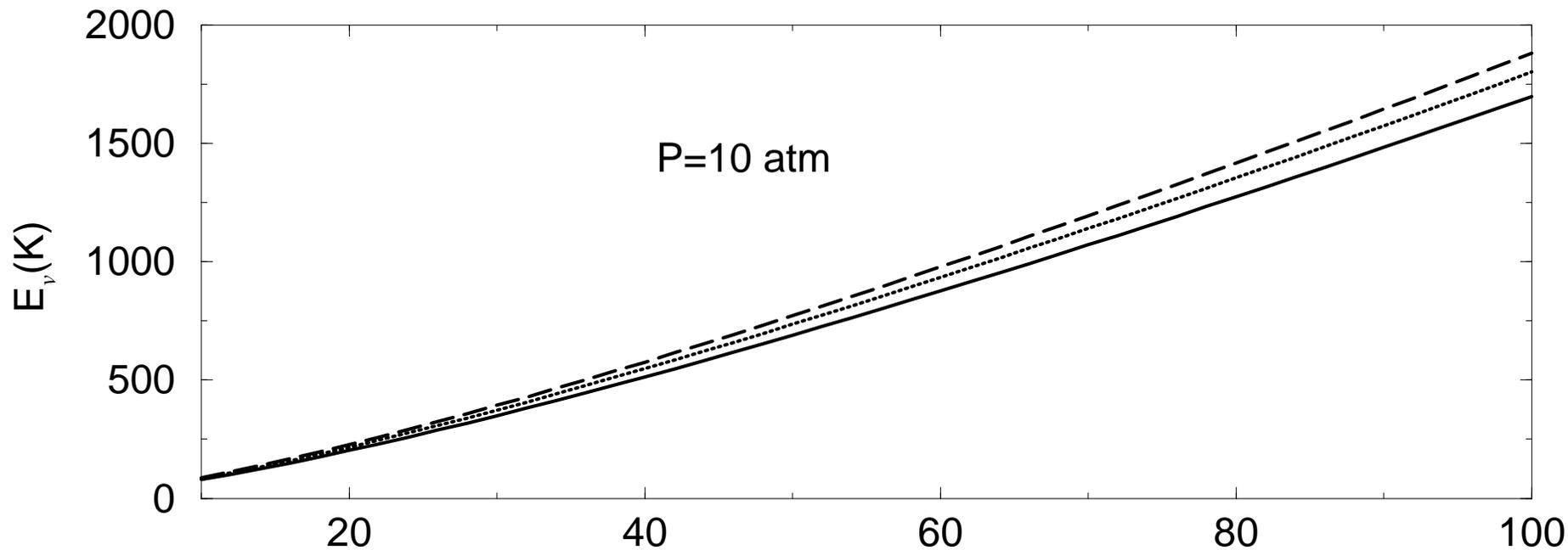
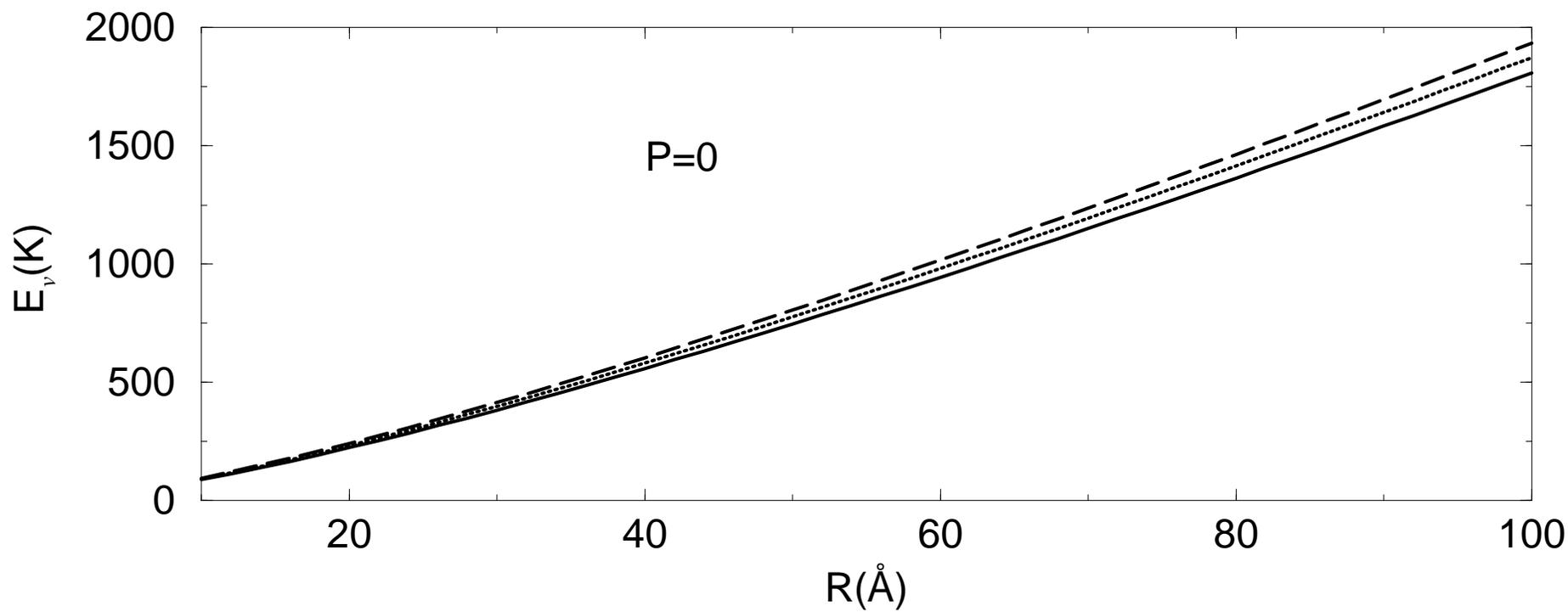
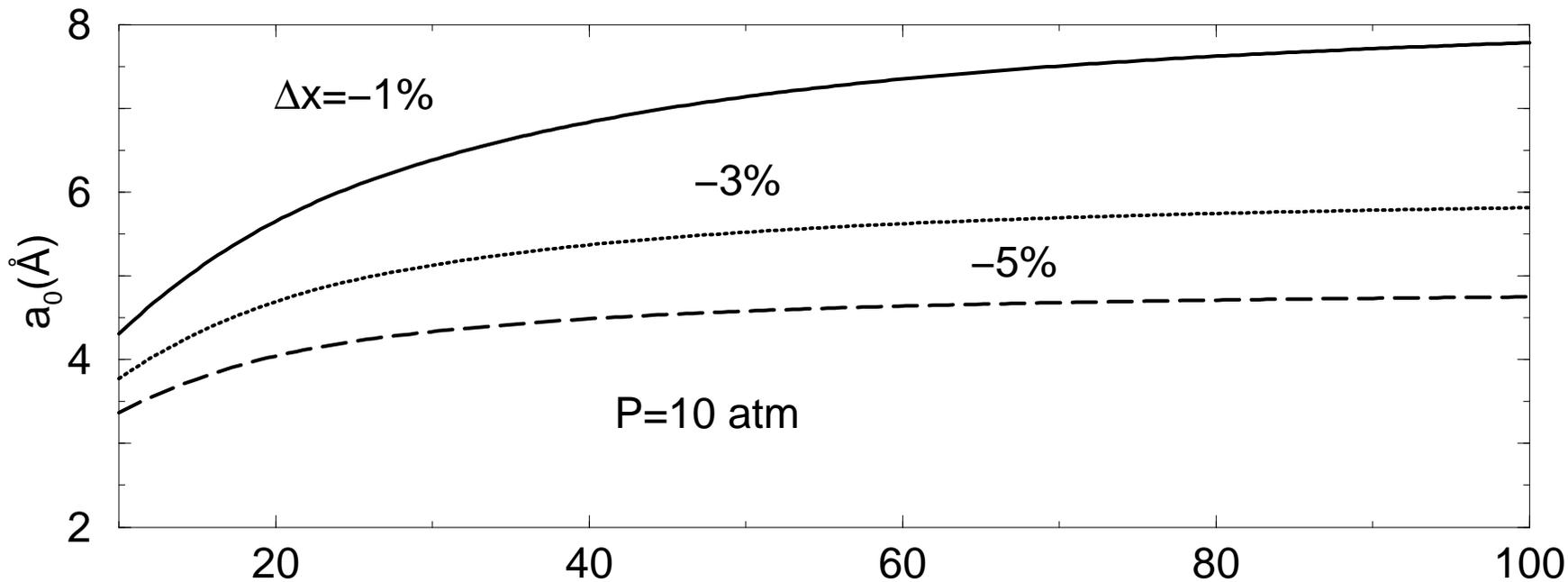
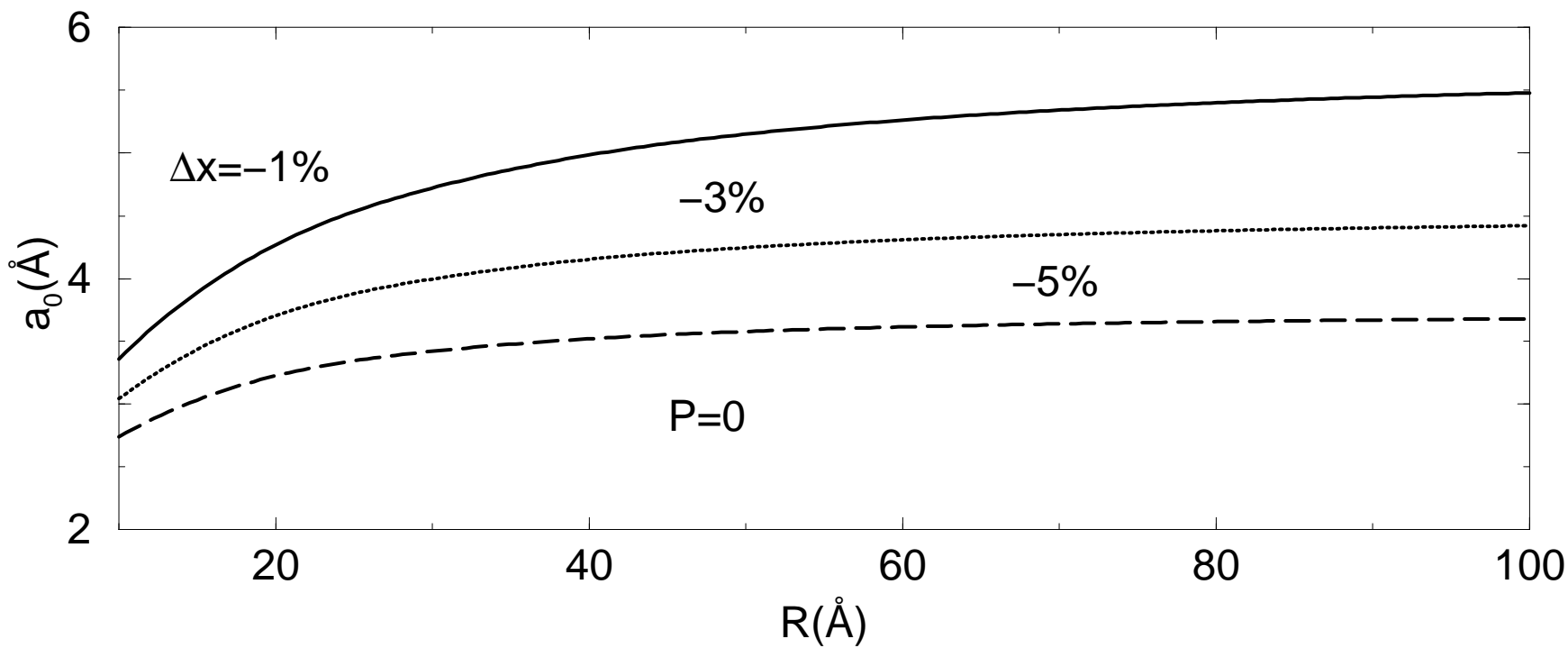
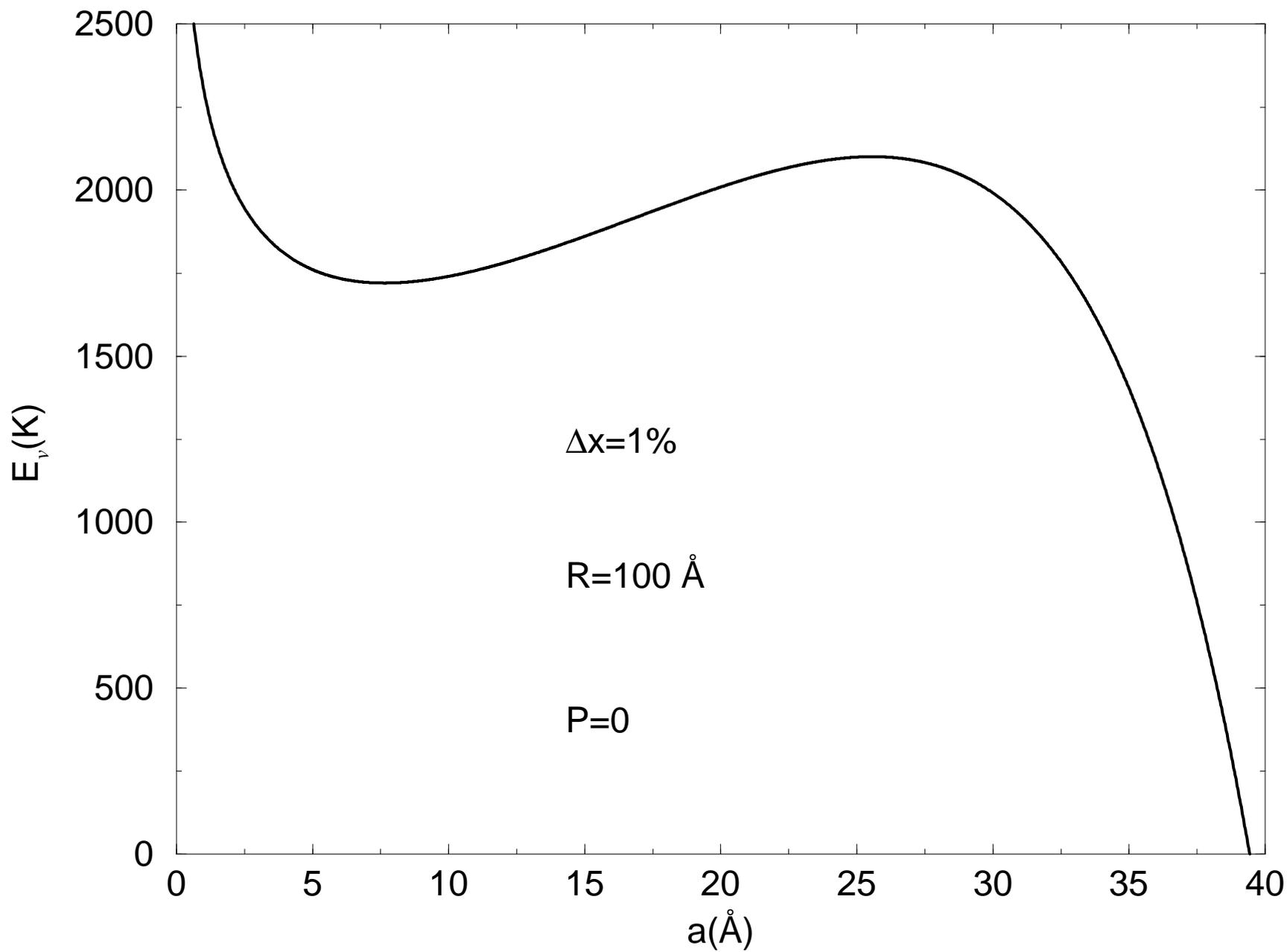
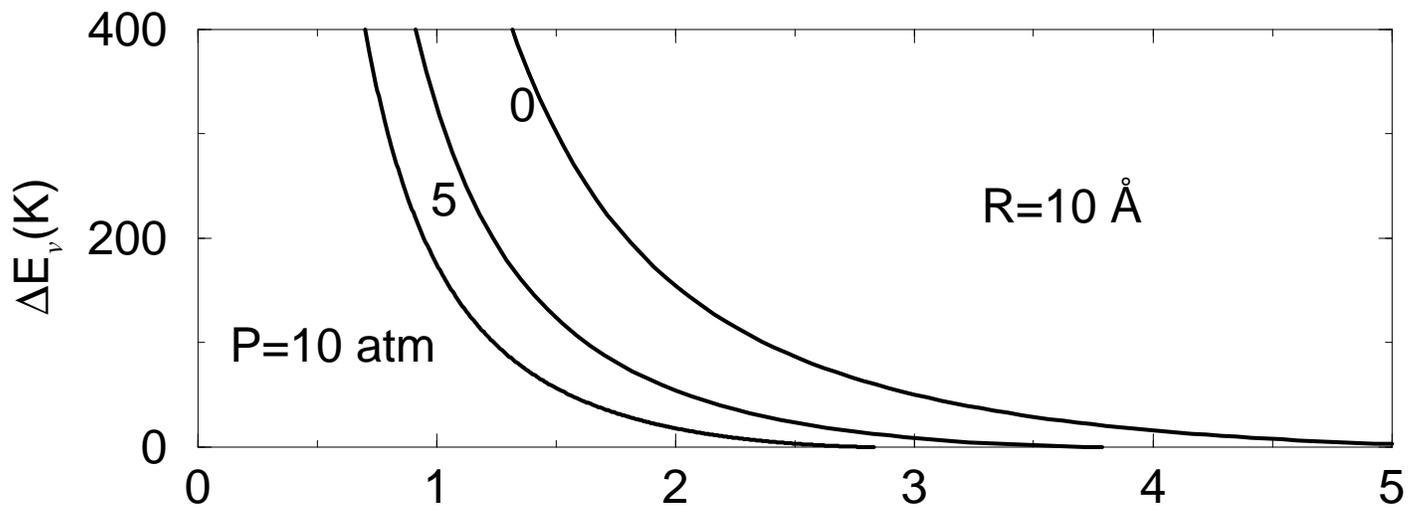
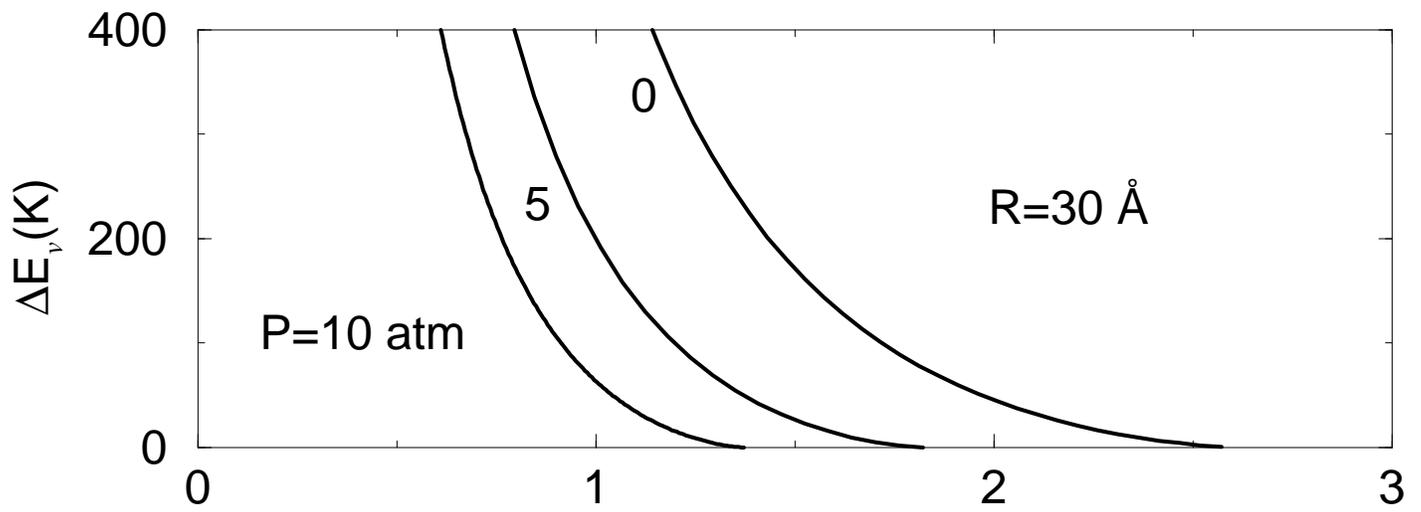
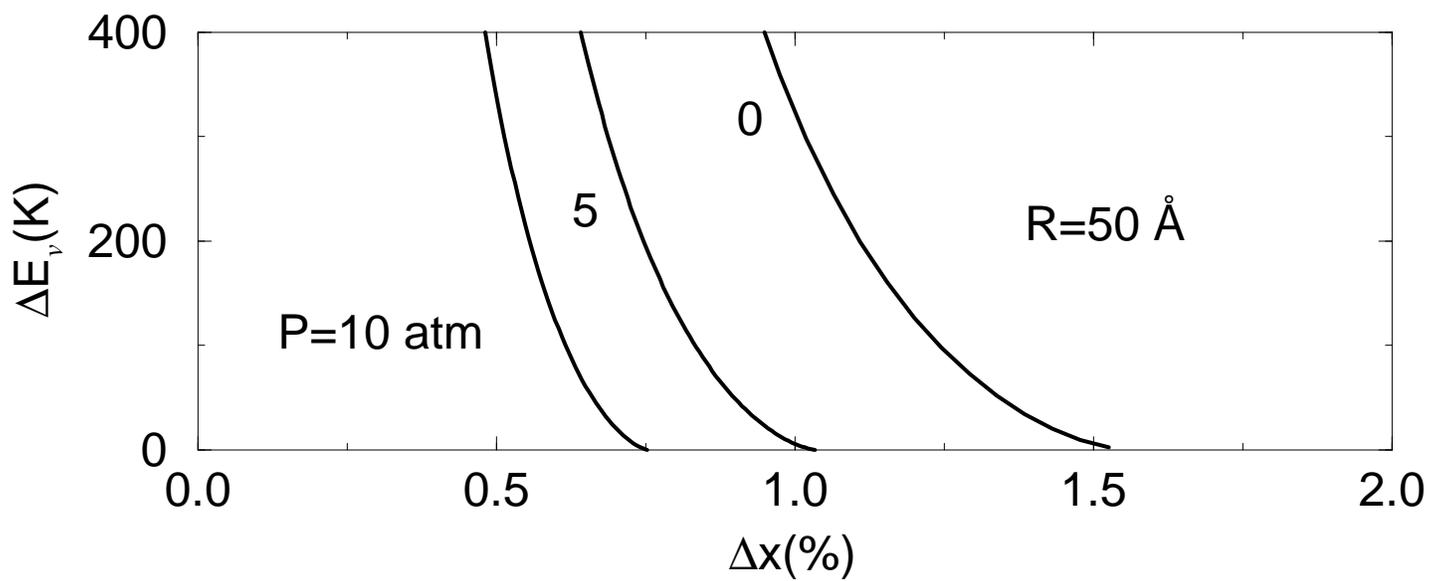
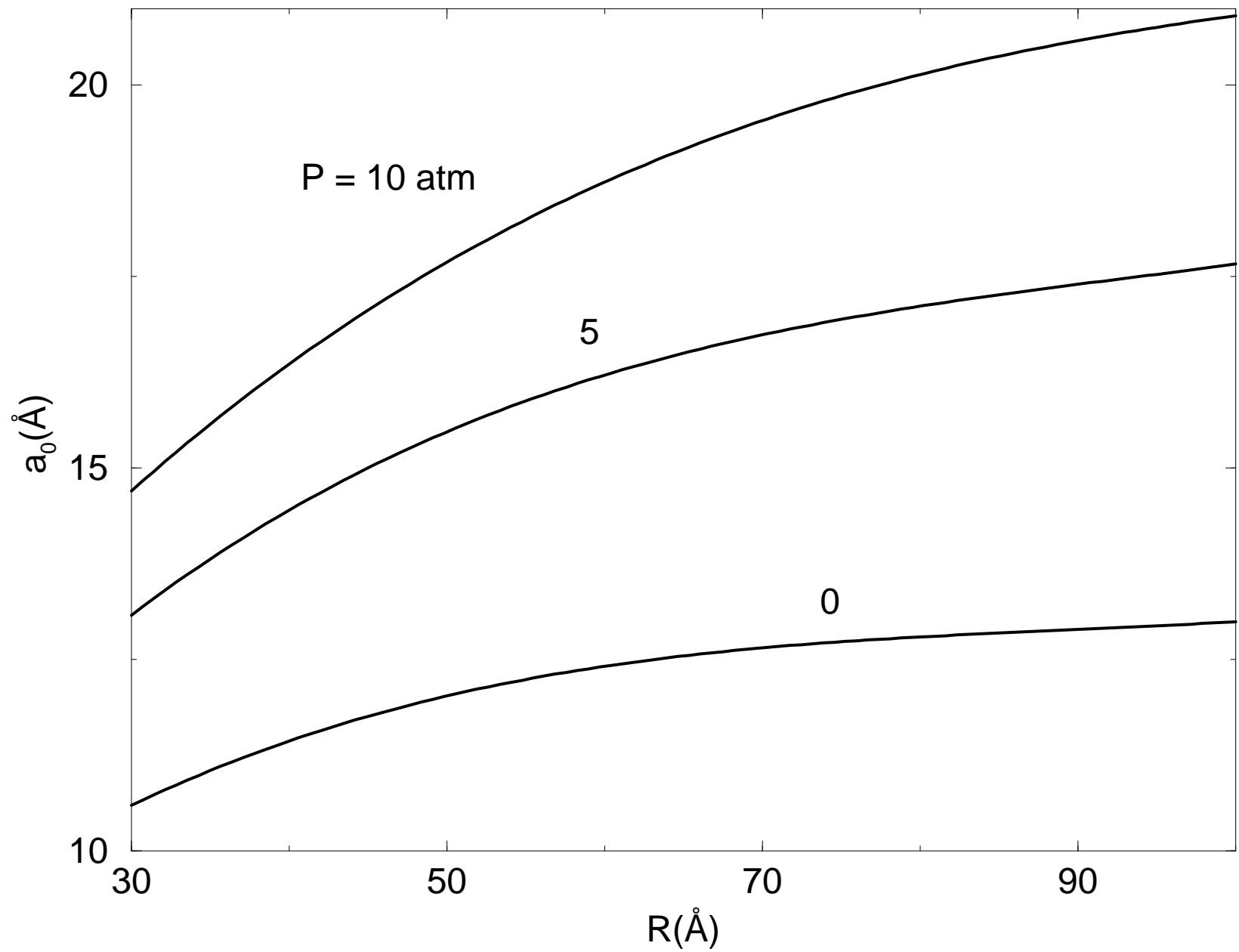
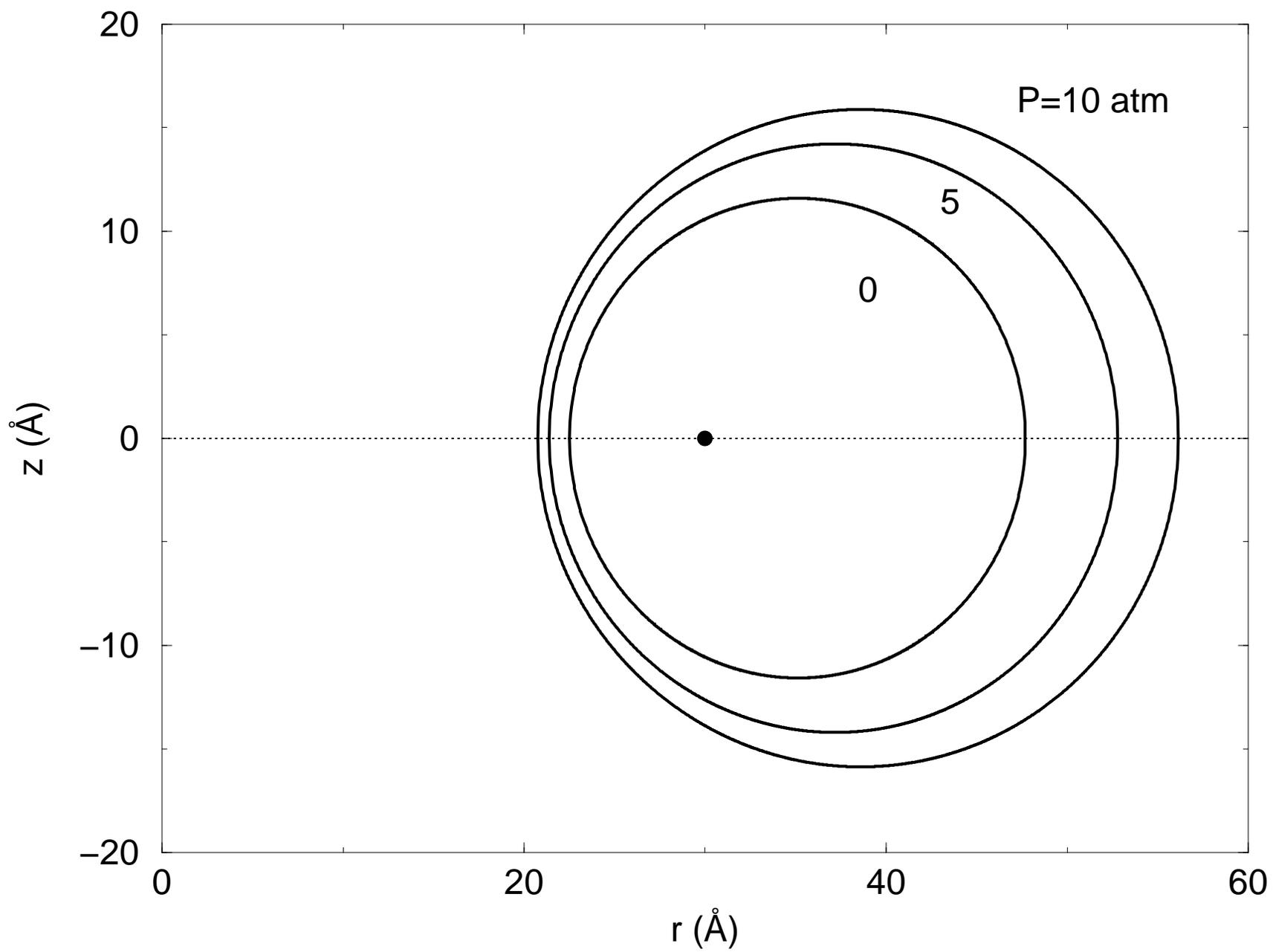
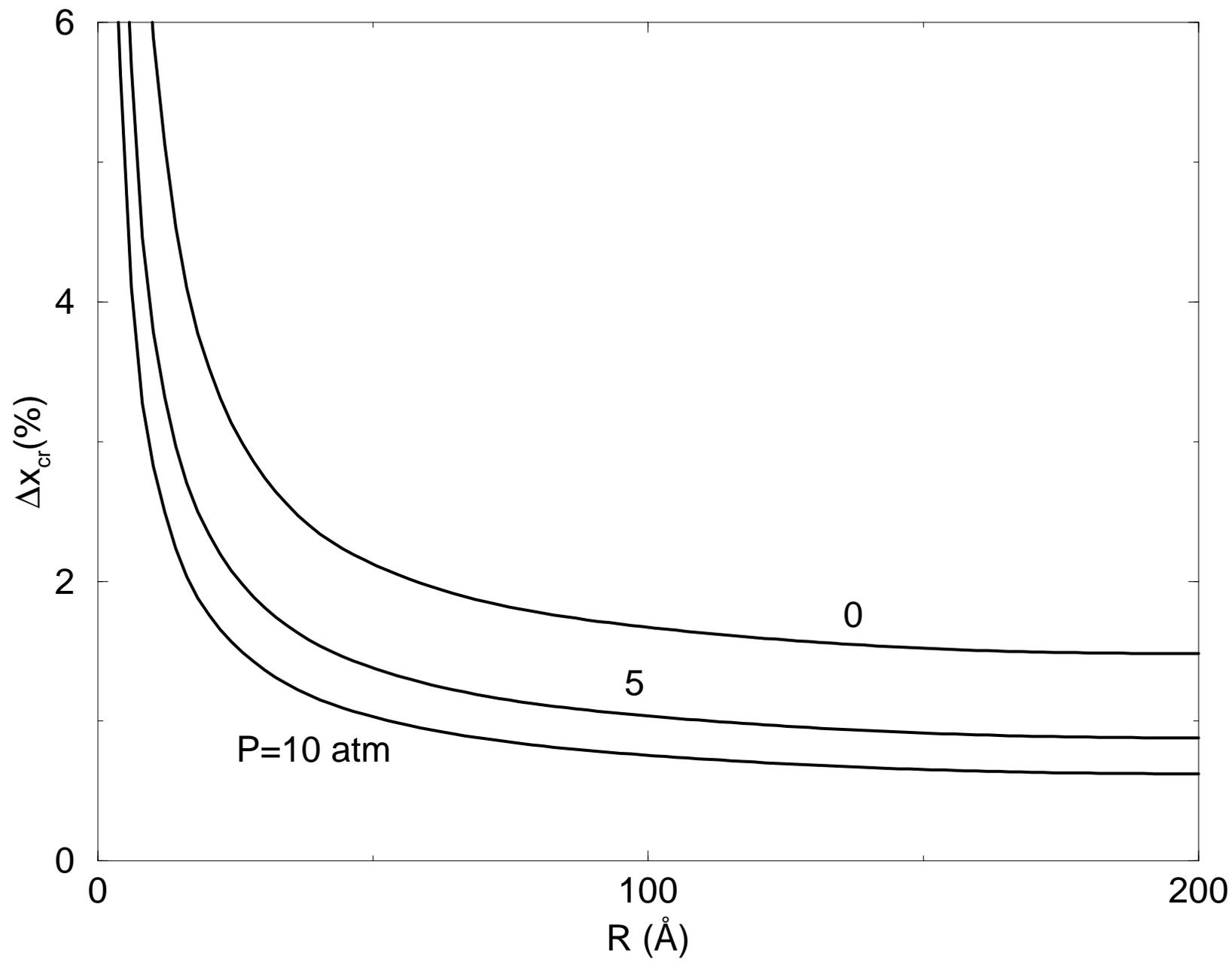
\end{document}